# TOWARD DEVELOPING IMPROVED
# MANDARINO COMPATIBILITY INDICES ($K_p/K_c$)

Stephen A. Langford * Ph.D.

* Langford Consulting, 805 N. Main St., Danielson, CT 06239

**ABSTRACT**

The specific-refractivity (K) model for a whole-rock powder is presented. Successful, innovative completion of this reconnaissance suggests that ignoring creation of statistically-significant specific refractivities ($K_p$), when describing mineral or rock samples, hampers the development of Science. Automation of all related technologies is suggested; as is the need for yet greater, more-widespread cooperation to work on standard samples, in order both to facilitate inter-laboratory instrument calibrations and to improve discrimination among samples.

**BACKGROUND**

The term "Specific Refractive Energy" was introduced by (Gladstone & Dale, 1863). Pointing to the preferred phrase "specific refractivity", Donnay, *et al*. (1980) nicely argued for dropping "energy" from the terminology and referenced (Gray, 1972); but the preferred term is also found in both (Gray, 1957) and (The Century Company, 1909).

Mandarino wrote a series of papers (1964, 1976, 1978, 1979, 1981, 2007). In (Mandarino, 1979), a distinction between $K_p$ (physical) and $K_c$ (chemical) specific refractivities was made. In this present paper, when without a subscript, K also means $K_p$. Although a bulk chemical analysis is available (Langford, 1972, p. 36), this present paper makes no attempt to sort chemistries to particular mineral species; that is to say, $K_c$ of the subject sample is completely ignored in this paper.

(Larsen, 1921) published what was to become (Larsen & Berman, 1934), in both of which Chapter 3 seems to be identical. Paradoxically (and perhaps to the detriment of encouraging people to work with $K_p$), $K_p$ values were not listed in the tabulations of either book. Sadly, the (Fleischer, *et al*., 1984) "revision" of (Larsen & Berman, 1934) completely redacted the Chapter 3 presentation of The Law of Gladstone and Dale! In (Mandarino, 2007, p. 1307), he wrote: "It is discouraging to see the lack of optical data in descriptions of new species". Happily, none of all that ever discouraged Mandarino from persevering in his amazingly gargantuan efforts to bring the chemical and physical aspects of that Law into a cohesive and functional methodology, as realized by his $K_p/K_c$ "compatibility index" (Mandarino, 1979, p. 71). The "The compatibility index using the Gladstone - Dale relationship … should also be calculated" advice by (Nickel & Grice, 1998) is also pertinent. However, due to such statements as

1) "Users of the Gladstone-Dale relationship should realize that wide differences between calculated and observed refractive indices and densities may be caused by the variation in k values" (Mandarino, 1964),
2) "During the latter half of the nineteenth century several other refractivity-density relations were proposed, but these had very restricted applicability and were little more than extrapolations from a limited number of experiments" (Kragh, 2018, p. 9; quoted by permission), and
3) "The practice of using the Gladstone-Dale relationship to minerals only gives an approximation because of the effects that different crystal systems have on the anisotropy of the crystal lattice and the resultant values of n (index of refraction)" http://webmineral.com/help/Gladstone-Dale.shtml;

and, due to the lack of prior work of the present kind: The purpose of this present paper is to encourage a general appreciation of precisions and accuracies available from statistically-significant refractive index (RI = n) data sets; and, to demonstrate how such work might elevate $K_p$ to the level of a statistic not to be ignored, but instead to be estimated with the greatest-possible precision and accuracy. In addition to the Lorenz-Lorenz work discussed by (Kragh, 2018), many other formulations suggesting improvements to the simplest K=Refractivity/Density approach (e.g., Sun, *et al*., 1940) have been encountered during literature reviews. To compare and contrast all such approaches is beyond the scope of this author.

Langford (1972, 1991, 2021a, 2021b) has worked extensively to investigate how much more might be learned from creating statistically-significant data sets, by studying RIs displayed by fragments of rock and mineral powders, via the double-variation, liquid-immersion methodology outlined by (Emmons, 1943, Ch. 5); but by bringing no fragment to any particular optical orientation during data collection. (Saylor, 1935) should be studied by any who would seriously emulate the techniques used to produce such results as those displayed below.

The copy of (Larsen & Berman, 1934) at hand – especially that part of Ch. 3 entitled "RELATION BETWEEN INDEX OF REFRACTION, DENSITY, AND CHEMICAL COMPOSITION" – has been copiously filled with marginalia since March of 1978, though the book was probably purchased circa 1969. While developing and applying the techniques more-fully detailed in the aforementioned prior papers, the thought of exploring how to map $K_p$ via some treatment of the data < https://tinyurl.com/v9u4hkf > was kept in mind.

**THE EXPERIMENT**

Abrupt transitions of RI-probability-levels are experimentally well defined for reasons explained by both (Saylor, 1935) and (Lipson & Lipson, 1969, p. 310. A complex Density model reflecting such complications was created by calculating the $K_p$ quotient of dividend mass (m) of (Langford, 2021b) over divisor E% (Langford, 2021a).

Intricate and complex modeling efforts produced both cross-hatched Kriging and Radial-Basis, Thin-Plate-Splining models, which were then combined into one model of Density. A model of Refractivity was also created, and the K=Refractivity/Density calculation was then performed, via matrix algebra in the Surfer® (Golden Software, Golden, Colorado) GRID|MATH routine.

Under the possibly incorrect assumptions that 1) specific refractivities are not additive in data from a whole rock powder RI study and 2) the $β'$ values derived from study of unoriented fragments (Langford, 1972) are close enough to those $(α+β+γ)/3$ values, for present purposes: The data of Table 1 (Appendix), calculated from (Fleischer, *et al*., 1984), provided the ability to calibrate those Density results by re-scaling them between the 0.18565217391304 minimum and the 0.21566137566138 maximum that reflect values related to the various mineral, solid-solution phases previously identified in the sample. Notably, prior work by (Jaffe, 1996; Fig. 11.2, p. 123) resulted in an interesting straight-line fit between measured densities and those calculated as (n-1)/K [i.e., (RI-1)/$K_p$].

## RESULTS

Results for the phase of ongoing work reported in this paper are best summarized by Figure 1.

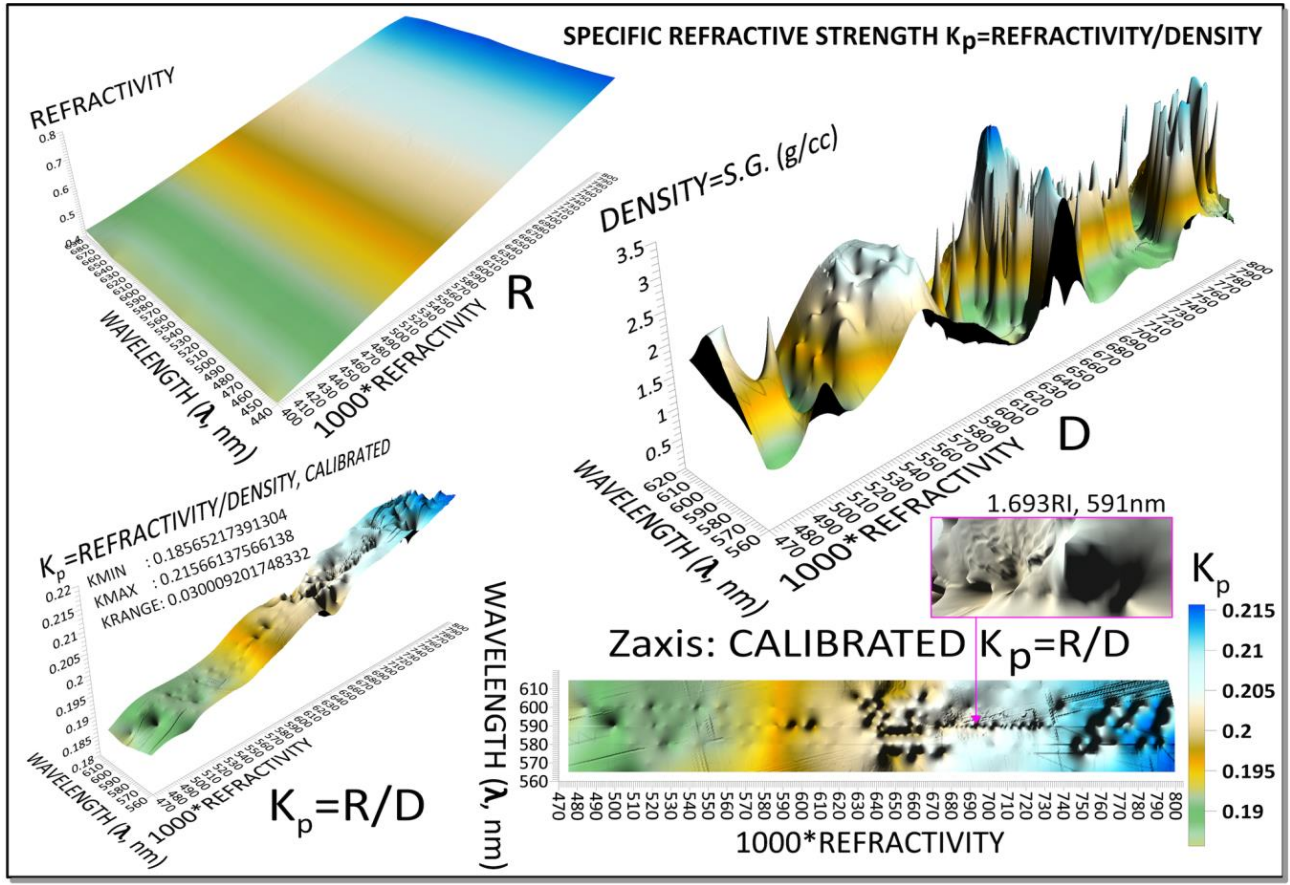

Fig. 1. Models of 1) Refractivity over most of the Emmons Surface within which data were taken during this study, 2) Densities estimated between 565 nm and 615 nm, and 3) $K_p$ values calculated over the same "Fraunhofer D Band" (ranging from 565 nm to 615 nm).

Were its most-meaningful profiles to be shown, they would be at 589.6 nm (where lots of D1 data were obtained via a dichroic filter); or, at the 589.3 nm "D line" (to which standard wavelength average of D1=589.6 nm and D2=589.0 nm Optical Mineralogists usually report RIs). Prior work (Langford, 2021[a,b]) shows results over the visible spectrum, but this paper is restricted to the region most pertinent to $K_p$ calculations as formulated in both (Larsen, 1921) and (Larsen & Berman, 1934) for the 589.3 nm Fraunhofer D-line data.

The (Fleischer, *et al*., 1984) redaction of what, in both (Larsen, 1921) and (Larsen & Berman, 1934), had been Chapter 3, "Some statistics on the optical properties of minerals"; at first seemed to suggest that the Law of Gladstone and Dale had been found to be useless! Though there are plethora references to "specific"; not one "energy", "refractivity" or "specific refractive" string is to be found in (Fleischer, *et al*., 1984), who offer (in part):

"The present edition, in holding to the objective of mineral identifications, omits discussion of statistics of mineral optics, as covered in Chapter 3 of the 1934 edition…. The Commission on New Minerals and Mineral Names, International Mineralogical Association (American Mineralogist, v. 67, p. 191-192), states for new minerals, 'It is recommended that the relationship between chemical composition, density, and refractive indices, be checked by the Gladstone-Dale rule.'"

## CONCLUSIONS

Stating that others recommend that chemistries, densities, and RIs be checked by the GD rule is not really stressing agreement with such advice. In this author's considered opinion, a revision of (Fleischer, *et al.*, 1984) is advisable. It would reinstate Chapter 3 (perhaps with appropriate updates) and list $K_p$ values for every mineral; and also list $K_c$, whenever those data are available for the same samples. Malcolm E. Back (who worked intimately with Joe Mandarino as a graduate student and who continues work relating $K_c$ to $K_p$) in recent personal communications has convinced this present author that at least some people have not lost sight of the importance of the Gladstone and Dale Law. This reinforces the suggestion that a reinstatement of Chapter 3 be made, in a revision of (Fleischer, *et al.*, 1984) wherein (for all appropriate tables) the empty spaces on the right side of the "Hardness, specific gravity, and fusibility" columns would be used to list all $K_p$ and relevant $K_c$ values. Future editions could include emboldened, $K_p$ data derived from statistically-significant data sets.

Improvements to methodologies more precisely and accurately to map RIs displayed by rock- and mineral-separate powders, via creation of statistically-significant RI data sets; together with the mapping of chemistries for the same standard samples (with similarly rigorous statistical approaches); might go a long way toward eliminating most of the scatter still to be found in such excellent work as Jaffe's (op. cit. Figure 11.2); and, toward facilitating revisions to petrographic, petrological, "mineralographic", and mineralogical classifications and nomenclatures (e.g. Mills, et. al.; 2009). This is far from a new cry: "Such evaluation should have been commonplace a generation ago" (Fairbairn, *et al.*, 1951); "Invariably, the organization that prepares a proposed standard cannot afford the time and effort to make all determinations necessary for standardization, and it must depend on the generosity of cooperating laboratories. Despite the large size of our organization, we must also depend on such assistance" (Flanagan, 1976). (Flanagan, 1986; p. 41) says that the Geostandards Newsletter was first issued in January 1977. The present author is both humbled and delighted to learn < https://tinyurl.com/p2yr346w > that in 2004 the journal was renamed *Geostandards and Geoanalytical Research*; and flourishes < https://tinyurl.com/rajnt5nk >. However, no references pertinent either to the precision and accuracy of refractive-index data sets or to Gladstone and Dale specific refractivities are yet to be found in Web searches of that journal. Pointers to such articles would be appreciated.

Automated data collection should be developed for all technologies related to specific refractivities, so as to create a truly practical, production-line system for coordinated physical and chemical descriptions and classifications applicable to all kinds of liquids and solids. Until such developments are realized and applied to a sufficient number of sample standards; and until more people are willing and able to cooperate more effectively on studies of whole-rock samples, as well as on carefully chosen particles for closest study by such as spindle-stage Optical Microscopy and Electron Microprobing; precisions and accuracies of compatibility indexes will continue to suffer; as will sample descriptions and inter-sample discriminations. The effects of rare-earth elements on $K_p$ values, for instance, might prove to be profound.

It has proven to be practicable to calculate a useful estimation of sample Densities by dividing the m Model of (Langford, 2021b) by a sufficiently well-determined E% Model (Langford, 2021a); then using that Density Model as the divisor into a well-determined Refractivity Model; in order to estimate $K_p$ values that were calibrated against calculations based upon trustworthy data in the literature (Fleischer, *et al*., 1984). This is believed to be the first such work done on any rock sample, permitting consideration of its $K_p$ gestalt.

In a personal communication, Malcolm Back wrote: "I think this [appearance of optical data in descriptions of new mineral species –SL] has improved somewhat in recent times and as I said the GD compatibility of Mandarino is still used today and is very often quoted in new mineral descriptions". This gives the present author reason to hope that this paper will be read by some who will recognize its significance and will rally to the cause of improving $K_p$ estimates; and, thereby, Mandarino compatibility-index ($K_p/K_c$) precisions and accuracies.

## WORK LOGS

Logs of all work done since 13 February 2021 are available at < https://tinyurl.com/f68djxfh >, which URL includes many statistical reports from regressions, model formulas, and model images; as well as developments of thoughts and adaptations of modeling techniques to purposes at hand. Even some Surfer® (.SRF-extension) models are included there, which those who have recent Surfer® versions can examine closely or "fly" against a simulated sky or other chosen background.

## ACKNOWLEDGMENTS

This paper has benefited significantly from Malcolm E. "Mal" Back's suggestions.

8Langford, S.A. (2021b) A Portion of Vacuum-Structure mass, energy, and weight. Canadian Journal of Pure and Applied Sciences. 15 (1):5169-5178.

Larsen, E.S. (1921) The Microscopic Determination of the Nonopaque Minerals. United States Geological Survey Bulletin 679, United States Government Printing Office, Washington, 289 pp.

Larsen, E.S., & Berman, H. (1934) The Microscopic Determination of the Nonopaque Minerals. United States Geological Survey Bulletin 848, United States Government Printing Office, Washington, 266 pp.

Lipson, S.G. & Lipson, H. (1969) Optical Physics. Cambridge University Press, 494 pp.

Mandarino, J.A. (1964) Critical analysis of the Gladstone-Dale Rule and its constants The Canadian Mineralogist 8, 137

Mandarino, J.A. (1976) The Gladstone-Dale Relationship - Part I: Derivation of new constants. The Canadian Mineralogist 14, 498 502.

Mandarino, J.A. (1978) The Gladstone-Dale relationship. Part II: Trends among constants. The Canadian Mineralogist 16, 169-174.

Mandarino, J.A. (1979) The Gladstone-Dale relationship. Part III: Some general Application. The Canadian Mineralogist 17, 71-76.

Mandarino, J.A. (1981) The Gladstone-Dale Relationship. IV. The compatibility concept and its application. The Canadian Mineralogist 19, 441-450.

Mandarino, J.A. (2007) The Gladstone Dale Compatibility of Minerals and its Use in Selecting Mineral Species for Further Study. The Canadian Mineralogist 45 (5), 1307 1324.

Mills, S.J., Hatert, F., Nickel, E.H. & Ferra, G. (2009) The standardisation of mineral group hierarchies: Application to recent nomenclature proposals. European Journal of Mineralogy, 21 (5), 1073 1080.

Nickel, E.H. & Grice, J.D. (1998) The IMA Commission on New Minerals and Mineral Names: Procedures and Guidelines on Mineral Nomenclature. The Canadian Mineralogist, 36 (3), 913 926.

Saylor, C.P. (1935) Accuracy of microscopical methods for determining refractive index by immersion. Part of Journal of Research of the National Bureau of Standards 15, 277-294.

Sun, K.-H., Safford, K.W., & Silverman, A. (1940) Review of the relation of density and refractive index to the composition of glass: II*. Journal of the American Ceramic Society, 23 (12), 343 354. doi:10.1111/j.1151-2916.1940.tb16034.x

**APPENDIX**

| | Mineral Species | α | β | γ | (α+β+γ)/3 | Refractivity | S.G. | Kcalculated | Bull1627 pp. |
|---|---|---|---|---|---|---|---|---|---|
| 1 | Tridymite | 1.482 | 1.484 | 1.486 | 1.4840 | 0.4840 | 2.25 | 0.2151 | 128 |
| 2 | Sanidine | 1.52 | 1.525 | 1.525 | 1.5233 | 0.5233 | 2.56 | 0.2044 | 219 |
| 3 | Sanidine | 1.523 | 1.529 | 1.53 | 1.5273 | 0.5273 | 2.56 | 0.2060 | 220 |
| 4 | Orthoclase | 1.519 | 1.524 | 1.525 | 1.5227 | 0.5227 | 2.57 | 0.2034 | 219 |
| 5 | Albite | 1.528 | 1.532 | 1.539 | 1.5330 | 0.5330 | 2.61 | 0.2042 | 136 |
| 6 | Albite | 1.527 | 1.534 | 1.535 | 1.5320 | 0.5320 | 2.6 | 0.2046 | 221 |
| 7 | Oligoclase | 1.536 | 1.541 | 1.546 | 1.5410 | 0.5410 | 2.64 | 0.2049 | 138 |
| 8 | Oligoclase | 1.536 | 1.541 | 1.546 | 1.5410 | 0.5410 | 2.64 | 0.2049 | 223 |
| 9 | Oligoclase | 1.541 | 1.546 | 1.55 | 1.5457 | 0.5457 | 2.65 | 0.2059 | 224 |
| 10 | Andesine | 1.546 | 1.55 | 1.554 | 1.5500 | 0.5500 | 2.67 | 0.2060 | 140 |
| 11 | Andesine | 1.546 | 1.55 | 1.554 | 1.5500 | 0.5500 | 2.65 | 0.2075 | 225 |
| 12 | Andesine | 1.549 | 1.552 | 1.556 | 1.5523 | 0.5523 | 2.66 | 0.2076 | 226 |
| 13 | Labradorite | 1.555 | 1.558 | 1.562 | 1.5583 | 0.5583 | 2.69 | 0.2076 | 142 |
| 14 | Labradorite | 1.558 | 1.562 | 1.566 | 1.5620 | 0.5620 | 2.69 | 0.2089 | 143 |
| 15 | Bytownite | 1.563 | 1.568 | 1.573 | 1.5680 | 0.5680 | 2.71 | 0.2096 | 143 |
| 16 | Bytownite | 1.565 | 1.569 | 1.574 | 1.5693 | 0.5693 | 2.72 | 0.2093 | 144 |
| 17 | Bytownite | 1.563 | 1.568 | 1.573 | 1.5680 | 0.5680 | 2.71 | 0.2096 | 230 |
| 18 | Bytownite | 1.565 | 1.569 | 1.574 | 1.5693 | 0.5693 | 2.72 | 0.2093 | 231 |
| 19 | Bytownite | 1.571 | 1.577 | 1.583 | 1.5770 | 0.5770 | 2.74 | 0.2106 | 233 |
| 20 | Anorthite | 1.575 | 1.584 | 1.589 | 1.5827 | 0.5827 | 2.76 | 0.2111 | 236 |
| 21 | Ol Forsterite | 1.653 | 1.664 | 1.686 | 1.6677 | 0.6677 | 3.35 | 0.1993 | 165 |
| 22 | Ol Chrysolite | 1.661 | 1.68 | 1.697 | 1.6793 | 0.6793 | 3.45 | 0.1969 | 169 |
| 23 | Ol Chrysolite | 1.661 | 1.68 | 1.697 | 1.6793 | 0.6793 | 3.15 | 0.2157 | 267 |
| 24 | Ol Chrysolite | 1.683 | 1.704 | 1.722 | 1.7030 | 0.7030 | 3.53 | 0.1992 | 275 |
| 25 | Ol Hyalosiderite | 1.71 | 1.733 | 1.75 | 1.7310 | 0.7310 | 3.69 | 0.1981 | 282 |
| 26 | Ol Tephroite | 1.75 | 1.766 | 1.779 | 1.7650 | 0.7650 | 3.87 | 0.1977 | 287 |
| 27 | Ol Hortonolite | 1.742 | 1.77 | 1.786 | 1.7660 | 0.7660 | 3.9 | 0.1964 | 289 |
| 28 | Ol Hortonolite | 1.758 | 1.786 | 1.804 | 1.7827 | 0.7827 | 4.1 | 0.1909 | 291 |
| 29 | Ol Ferrohorton... | 1.777 | 1.818 | 1.828 | 1.8077 | 0.8077 | 4.21 | 0.1918 | 295 |
| 30 | Ol Fayalite | 1.803 | 1.843 | 1.851 | 1.8323 | 0.8323 | 4.3 | 0.1936 | 297 |
| 31 | Ol Fayalite | 1.816 | 1.85 | 1.863 | 1.8430 | 0.8430 | 4.36 | 0.1933 | 298 |
| 32 | Ol Liebenbergite | 1.82 | 1.854 | 1.888 | 1.8540 | 0.8540 | 4.6 | 0.1857 | 298 |
| 33 | Augite | 1.714 | 1.723 | 1.774 | 1.7370 | 0.7370 | 3.55 | 0.2076 | 179 |
| 34 | Augite | 1.725 | 1.73 | 1.75 | 1.7350 | 0.7350 | 3.42 | 0.2149 | 181 |
| 35 | Pigeonite | 1.684 | 1.684 | 1.707 | 1.6917 | 0.6917 | 3.4 | 0.2034 | 170 |
| 36 | Pigeonite | 1.696 | 1.698 | 1.721 | 1.7050 | 0.7050 | 3.38 | 0.2086 | 173 |
| 37 | Pigeonite | 1.714 | 1.714 | 1.742 | 1.7233 | 0.7233 | 3.44 | 0.2103 | 176 |

Table 1. $K_p$ values calculated from USGS Bull. 1627 (Fleischer, *et al.*, 1984).